\documentclass[reprint, amsmath,amssymb, aps,prc, superscriptaddress, showpacs]{revtex4-1}
\usepackage{graphicx}
\usepackage{dcolumn}
\usepackage{bm}

\begin{document}
\preprint{aps}
\title{Pre-equilibrium  mechanisms in the $^{93}$Nb($\vec{p}$,$\alpha $) inclusive reaction at incident energies from 65 to 160 MeV}%

\author{S.~S.~Dimitrova}
\email[]{sevdim@inrne.bas.bg}
\affiliation{ Institute for Nuclear Research and Nuclear Energy, Bulgarian Academy of Sciences, 1784 Sofia, Bulgaria}
\affiliation{ Joint Institute for Nuclear Research, 141980 Dubna, Russia}
\author{A.~A.~Cowley}
\email[]{aac@sun.ac.za}
\affiliation{ Department of Physics, Stellenbosch University, Private Bag X1, Matieland, 7602, South Africa}
\affiliation{ iThemba Laboratory for Accelerator Based Sciences, P O Box 722, Somerset West 7129, South Africa}
\author{E.~V.~Zemlyanaya}
\affiliation{ Joint Institute for Nuclear Research, 141980 Dubna, Russia}
\author{K.~V.~Lukyanov}
\affiliation{ Joint Institute for Nuclear Research, 141980 Dubna, Russia}

\begin{abstract}
The reaction mechanism of pre-equilibrium proton-induced $\alpha $-particle emission from $^{93}$Nb at an incident energy
of 100 MeV was investigated with polarized projectiles. A formalism based on the statistical multistep
direct emission model of  Feshbach, Kerman and Koonin was found to give
a reasonably good reproduction of cross section and analyzing power angular distributions at various emission energies. Existing experimental distributions for the  same reaction at an incident energy of 65 MeV
were also analyzed with the same model. The incident-energy variation from 65 MeV up to 160 MeV was found to be consistent with the predictions of the basic model.
However, whereas knockout of an $\alpha $ cluster is the dominant reaction mechanism in the final stage at the lowest- and highest incident energies, at 100 MeV a pickup
process competes with comparable intensity in yield.

\end{abstract}


\maketitle

\section{Introduction}

Angular-  and energy distributions of  nucleons emitted in proton-induced pre-equilibrium reactions \cite{Gad91} in the incident energy range up to
200 MeV are described well in terms of several related quantum-mechanical formulations \cite{Kon93,Gad91}. Of the available models, the statistical multistep
direct emission (SMDE) theory of Feshbach, Kerman and Koonin (FKK) \cite{Fes80} has been extensively and successfully compared with experimental results over a
large target mass and incident energy range \cite{Gad91,Ric96,Ric94,Ric92,Cow91}.

Although the emission of composite particles in pre-equilibrium reactions, such as $^{3}$He and $\alpha $ particles, could
be a more complicated process than the emission of nucleons, it is nevertheless reasonable to expect that
the  relevant reaction mechanism should be an intrinsic part of the basic
process described, for example, by the FKK theory.

In earlier work on proton-induced emission of $^{3}$He  into the continuum, we attempted to identify the simplest dominant reaction process in the
incident energy range below 200 MeV \cite{Cow97,Cow00,Cow07,Cow12}. Because analyzing power angular distributions are more sensitive to details of the reaction mechanism than
those of the cross section, polarized projectiles proved to be especially valuable for these studies. It was found that the incident-energy evolution of the
characteristics of the analyzing power angular distributions is consistent \cite{Cow00,Cow07,Cow12} with a simple two-nucleon pickup process convoluted with the SMDE mechanism.
Unfortunately, the usefulness of the analyzing power to unravel details of the reaction mechanism diminishes towards the upper end
of the energy range. The reason is that the analyzing power tends to have lower values at higher incident energies, essentially
disappearing at 200 MeV \cite{Ren91} even at forward angles.
The observed quenching of the analyzing power as a function of increasing incident energy is understood \cite{Cow14} as an inherent feature of direct pickup.

The present study provides further insight into the reaction mechanism of emission of $\alpha $ particles. Of course, three-nucleon pickup as well as knockout
could both be important processes in proton-induced emission of $\alpha $ particles. However, in one of our early
investigations with unpolarized projectiles \cite{Cow96} at incident energies between 120 and 200 MeV
the experimental cross section angular distributions of the reaction $^{59}$Co($p$,$\alpha $) at various emission energies were
fairly reliably reproduced by invoking only a knockout mechanism in the theoretical analysis. Also, in our
recent paper \cite{SCZ14} on the reaction
$^{93}$Nb($\vec{p}$,$\alpha $) at an incident energy of 160 MeV, we found that best agreement of the theoretical predictions with experimental cross section and analyzing power
angular distributions is obtained if knockout is assumed as the dominant mechanism. While keeping in mind these earlier results, in the present study we again consider
the possible participation of both mechanisms -- pickup as well as knockout -- in the pre-equilibrium ($p$,$\alpha $) reaction at the lower incident energies explored now.

The motivation for the present work is to investigate the incident-energy dependence of the $^{93}$Nb($\vec{p}$,$\alpha $) reaction to lower values
(down to 65 MeV) than explored in our previous investigation at 160 MeV. For this purpose we use new experimental data at 100 MeV together with existing published
angular distributions at an incident energy of 65 MeV \cite{Sakai80}. We assume implicitly that the reaction mechanism for the target nucleus $^{93}$Nb is
representative of nuclei in general.
Clearly, trivial differences, which relate to structure details of a particular nuclear species, should be observable.

We now find, as in our earlier work and as would be expected, that the extended FKK theory gives a very good reproduction of the cross section and analyzing power angular distributions
for the $^{93}$Nb($\vec{p}$,$\alpha $) reaction at 100 MeV as well as at 65 MeV incident energy.  However, an unexpected and interesting
feature of the  new investigation is that, at an incident energy of 100 MeV, a pickup mechanism now competes strongly with a knockout process.
Evidently, this is in strong contrast with the trend at both higher (160 MeV \cite{SCZ14}) and lower (65 MeV; also from this work) incident energies where knockout appears to be overwhelmingly dominant.

This paper has the following structure: In Sec. II the experimental technique at an incident energy of 100 MeV is described. This is followed with a summary of the
theoretical ideas in Sec. III. In Sec. IV the results are shown and discussed. Finally, in Sec. V a summary and conclusions are presented.

In this paper we often use, for example, the notation ($p$,$\alpha $) instead of ($\vec{p}$,$\alpha $) which is appropriate. As we refer mostly to reactions induced by polarized projectiles,
the meaning should be clear from the context.

\section{Experimental procedure}

The reaction $^{93}$Nb($\vec{p}$,$\alpha $) at an incident energy of 100 $\pm $ 0.5 MeV was measured
at iThemba LABS (at the time known as the National Accelerator Centre) in Faure, South Africa.
A description of the facility is available in Ref. \cite{Pil89}.

Cross sections and analyzing distributions were measured in the same experiment \cite{Cow00} as those for $^{93}$Nb($\vec{p}$,$^{3}$He), but
the present data were extracted at a much later date from the event-by-event records stored online.
Only a brief summary of the fairly standard experimental technique is provided here for ease of reference.

Two detector telescopes, each consisting of a 500-$\mu $m silicon surface-barrier detector followed by a
NaI(T$\ell $) crystal coupled to a phototube, were
positioned at symmetric angles on opposite sides of the incident beam in a 1.5-m diameter scattering chamber. The telescopes
were collimated to a solid angle acceptance of about 1.1 msr. The scattering-angle
positions were set to an accuracy of better than 0.2$^\circ $ with respect to the incident beam.

Two self-supporting targets of naturally occurring niobium (100$\% $ in the isotope $^{93}$Nb) of
thicknesses of approximately 1 and 5 mg/cm$^2$ were used. The main systematic uncertainty in the cross section
data -- about 8$\% $ -- originates from the
absolute value of the target thickness and its uniformity.

The incident proton beam was polarized to a nominal value of 80$\% $ perpendicular to the reaction plane,
and the direction of the polarization was switched at 5-s intervals during measurements. Variation between
the degree of polarization for the two directions was less than 10$\%$. These values were monitored regularly
by means of elastic scattering of the proton beam from a carbon target at a
scattering angle where the analyzing power is large and known accurately.

The use of detector telescopes positioned symmetrically with respect to the incident proton beam, together
with the switching of the polarization direction allows us to minimize systematic errors in the
analyzing power measurements. The vector analyzing power is calculated from the expression \cite{Sakai80},
which follows from the standard Basel-Madison conventions, as
\begin{equation}
A_{y}=\frac{L-R}{P(L+R)},
\end{equation}
with
\begin{equation}
L=\sqrt{L_{u}R_{d}},
\end{equation}
and
\begin{equation}
R=\sqrt{L_{d}R_{u}}.
\end{equation}
The average
polarization of the beam is $P$.  The summed counts in each detector for a given energy interval in the spectra are indicated by
$L$ (left) or $R$ (right), with subscripts which indicate the spin direction of the projectile as either up ($u$) or down ($d$). The convention
is as defined by a spectator facing along the momentum direction of the incident beam upstream from the target.  Comparison of this formulation of
analyzing power with the expression containing the specific values of
the two orientations of the polarization \cite{LLW82} indicates that a 10$\% $ difference affects the
measured value by only 1$\% $.

Energy calibrations of the Si detectors were based on measurements from a $^{228}$Th source, and those of the NaI crystals
were determined from proton scattering from a (CH)$_{n} $
target, adjusted for the difference in response of $\alpha $ particles. The overall accuracy of the emission-energy scale is better than 4$\%$.
Cross sections and analyzing powers were binned in 4 MeV wide energy intervals.

\section{Theoretical Analysis}

We consider the ($p$,$\alpha $) inclusive reactions at incident energy of 100~MeV and 65~MeV as pre-equilibrium reactions. As in our
previous studies on ($p$,$^{3}$He) processes \cite{Cow97,Cow00,Cow07, Cow12}, we assume that this type of reaction occurs in a series
of intranuclear $N$-$N$ steps
preceding a final process in which the $\alpha $ particle is emitted. The single step direct reaction can be a knockout of
an $\alpha$ cluster or a pickup of a triton. We will consider the contribution of both reaction mechanisms to the total
double-differential cross section and analyzing power for different energies of the $\alpha$ particle in the outgoing channel.

The theory applied to the ($p,\alpha $) reaction is based on the multistep
direct theory of Feshbach, Kerman and Koonin (FKK) \cite{Fes80}. The extension of the FKK theory from nucleon- to composite-particle emission
has been  presented often, and a recent description can be found in our previous paper \cite{SCZ14}.

The details of the methodology of the ($p$,$\alpha$) calculations are also described in Ref. \cite{SCZ14}, thus now we
will just briefly outline the main expressions. We will emphasize specific subtleties needed for the adequate description of the reactions considered.

\subsection{Differential cross sections}
The expression for the pre-equilibrium ($p,\alpha $) cross section is written as a sum of various steps as
\begin{eqnarray}
&&\hspace{-0.5cm}
\left( \frac{d^{2}\sigma }{d\Omega dE}\right)_{{p, \alpha}}^{\rm{total}} =\left( \frac{d^{2}\sigma }{d\Omega dE}
\right)_{{p, \alpha}} ^{\rm{1-step}}
\notag \\
&&\hspace{-0.5cm}+\sum\limits_{n=2}^{n_{\max }}\sum\limits_{m=n-1}^{n+1}\int
\frac{d\mathbf{k}_{1}}{(2\pi )^{3}}\int \frac{d\mathbf{k}_{2}}{(2\pi )^{3}}%
\ldots \int \frac{d\mathbf{k}_{n}}{(2\pi )^{3}}  \notag \\
&&\hspace{-0.5cm}\times \left( \frac{d^{2}\sigma (\mathbf{k}_{f},\mathbf{k}%
_{n})}{d\Omega _{f}dE_{f}}\right) \times \left( \frac{d^{2}\sigma (\mathbf{k}%
_{n},\mathbf{k}_{n-1})}{d\Omega _{n}dE_{n}}\right) \times \ldots  \notag \\
&&\hspace{-0.5cm}\times \left( \frac{d^{2}\sigma (\mathbf{k}_{2},\mathbf{k}%
_{1})}{d\Omega _{2}dE_{2}}\right) \times \left( \frac{d^{2}\sigma (\mathbf{k}%
_{1},\mathbf{k}_{i})}{d\Omega _{1}dE_{1}}\right) _{p,N},
\end{eqnarray}
\noindent where $\mathbf{k}_{i},\ \mathbf{k}_{n}$ and $\mathbf{k}_{f}$ are the momenta of the initial, $n^{th}$ and final steps.
The number of reaction steps is indicated with the symbol $n$, the maximum number of reaction steps is $n_{\rm{max}}$ and $m$ is the exit mode. Therefore, in the present application, the cross section
associated with $m$ corresponds to the emission of an $\alpha $ particle. All steps prior to the final emission are nucleon-nucleon collisions which
originate from the initial projectile-$N$ cross section $\left( \frac{d^{2}\sigma }{d\Omega _{1}dE_{1}}\right) _{p,N}$. Of course, the first term, which does not involve preceding nucleon collisions, is
given in terms of the distorted-wave Born approximation (DWBA) by

\begin{equation}
\left( \frac{d^{2}\sigma }{d\Omega dE}
\right)_{{p, \alpha}} ^{\rm{1-step}}\!\!\!=\sum_{N,L,J}{%
\frac{(2J+1)}{\Delta E}}{\frac{d\sigma ^{\rm{DW}}}{d\Omega }}(\theta ,N,L,J,E)\;%
\text{,}  \label{eq2}
\end{equation}

\noindent at scattering angle $\theta $, where the summation runs over the target states with single-particle energies within a small interval $(E-\Delta E/2,E+\Delta E/2)$ around the
excitation energy $E$ (in these particular calculations we adopted $\Delta E$=4~MeV to match the experimental energy bin). If the DWBA calculation is treated as a knockout,
quantum numbers $N,L$ and $J$ refer to the $\alpha $ cluster bound in the target, otherwise to those of the three-nucleon system which is picked up. The differential cross
sections $d\sigma ^{\rm{DW}}/{d\Omega }$ to particular $(N,L,J)$ states are calculated using the code \textsc{DWUCK4} \cite{KR93}.

A related formulation in terms of the DWBA holds for the initial projectile-$N$ interaction represented by $\left( \frac{d^{2}\sigma }{d\Omega _{1}dE_{1}}\right) _{p,N}$. This is provided in, for example, Ref. \cite{Cow91}.

The theoretical ($p,p^{\prime }$) and ($p,p^{\prime },p^{\prime \prime }$) double-differential cross section distributions which are required to calculate the contributions
of the second- and third-step processes were derived from Refs. \cite{Ric94,Cow96}. These cross section distributions
were extracted by means of a FKK multistep direct reaction theory, which reproduces experimental inclusive $(p,p^{\prime })$ quantities \cite{Ric94} on
target nuclei which are close to those needed for this work, and in an appropriate incident energy range. Interpolations and extrapolations in incident
energy and target mass were introduced to match the specific
requirements accurately.

Clearly, the formalism separates calculation of multistep processes, such as one-step $(p,\alpha )$, two-step $(p,p^{\prime },\alpha )$, and three-step $(p,p^{\prime },p^{\prime \prime },\alpha )$ reactions.
This can be expressed as
\begin{equation}
\frac{d^{2}\sigma }{d\Omega dE}=\left( \frac{d^{2}\sigma }{d\Omega dE}%
\right) ^{\rm{1-step}} \!\!+\left( \frac{d^{2}\sigma }{d\Omega dE}\right)
^{\rm{2-step}} \!\! + \, \cdots \, ,
\end{equation}

\noindent in which the relationship of the notation is linked clearly to the formulation given thus far.

In previous work
\cite{Cow00,Cow07}, intermediate steps which involve neutrons, such as $ (p,n,\alpha )$, were not explicitly taken into account because we assumed that different nucleons may be
treated on an equal footing in the multistep part of the reaction. This meant that a simple renormalization of the ($p,p^{\prime }$) and ($p,p^{\prime },p^{\prime \prime }$)
cross sections should be introduced to correct for the influence of the intermediate counterparts which involve neutrons.
In these present calculations we take into account explicitly the $ (p,n,\alpha )$ process by
assuming that $ d^{2}\sigma^{(p,n)}/{d\Omega dE} = d^{2}\sigma^{(p,p^{\prime})}/{d\Omega dE}$ and also the four possible combinations
of two-step intranuclear collisions $ (p,x,x ), x=n,p$ with $ d^{2}\sigma^{(p,x,x)}/{d\Omega dE} = d^{2}\sigma^{(p,p^{\prime},p^{\prime\prime})}/{d\Omega dE}$.

\subsection{Analyzing power distributions}
The extension of the FKK theory from cross sections to analyzing power is described by Bonetti \textit{et al.} \cite{Bon82}. The multistep expression for the analyzing power becomes

\begin{equation}
A_{\rm{multistep}}=\frac{A_{1}\left( \frac{d^{2}\sigma }{d\Omega dE}\right)
^{\rm{1-step}}\!\!+A_{2}\left( \frac{d^{2}\sigma }{d\Omega dE}\right)
^{\rm{2-step}}\!\!+\cdots}{\left( \frac{d^{2}\sigma }{d\Omega dE}\right)
^{\rm{1-step}}\!\!+\left( \frac{d^{2}\sigma }{d\Omega dE}\right)
^{\rm{2-step}}\!\!+\cdots},
\end{equation}

\noindent with $A_{i}$, $\{i=1,2,\ldots \}$ referring to analyzing powers for the successive multisteps.

\subsection{Multi-step contributions to the cross section and analyzing power}

\begin{figure}[htb]
\includegraphics[scale=0.48]{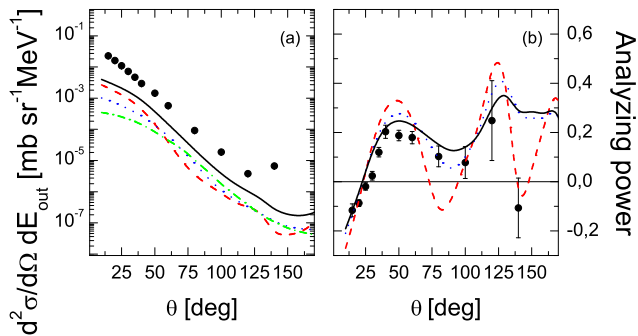}
\caption{(Color online) Double-differential cross section (a) and analyzing power (b) as a function of scattering angle $\protect\theta $ for the $^{93}$Nb($p$,$\alpha $)
reaction at an incident energy of 100~MeV and an $\alpha $-particle emission energy of 86~MeV. Results for only the pickup component of the reaction mechanism are used to display contributions of various steps.
Theoretical cross section calculations for one step ($---$), two steps
($\cdot \cdot \cdot \cdot \cdot $) and three steps
($-\cdot -\cdot -$) are shown, with the sums of the contributions plotted as continuous curves. The experimental analyzing power distribution is compared with theoretical calculations for a one-step reaction ($---$), a
one-step plus a two-step reaction ($\cdot \cdot \cdot \cdot \cdot $), and a one- plus two- plus three-step reaction (solid lines).}
\label{fig.1}
\end{figure}

In Fig. \ref{fig.1} the one-, two- and three-step contributions to the double-differential cross section and analyzing power as a function
of scattering angle $\protect\theta $ for the $^{93}$Nb($p$,$\alpha $) reaction at an incident energy of 100~MeV and an $\alpha $-particle emission
energy of 86~MeV are displayed. For the purpose of this illustration of the effect of contributions from various steps of the interaction, only the pickup component of the reaction mechanism is used.

For this particular case of energy transfer all three steps of the process contribute significantly to the double-differential
cross section and lead to a reduction of the
oscillatory behavior of the analyzing power associated with the first step. This effect can be understood qualitatively in terms of the formulation of the combined analyzing power of the contributing multisteps.

The theory predicts that the relative contribution of the
first-step reaction decreases as the emission energy drops, with higher steps becoming progressively more important towards lower emission energy. This is a general feature of
multistep calculations, as was also found in our previous work \cite{Cow00,Cow07,Cow97,Cow12,SCZ14}. Although the actual step which is dominant at a specific emission energy only
influences the shape of the cross section relatively slightly, an appreciable contribution of higher steps affects the analyzing power distribution profoundly. The trend is that
the analyzing power tends towards zero at lower emission energy where higher steps become more important.

\subsection{Optical potentials in the DWBA calculation}

As in our previous papers \cite{Cow12,SCZ14}, in the DWBA calculations we employ the microscopic optical potential
which takes into account the interaction between the
projectile and target, and between the ejectile and the heavy residual nucleus, respectively.
The theoretical approach of the microscopic optical potential is presented in Refs.~\cite{Khoa2000,Shukla2003,Lukyanov2004,Kostya2007} and successfully applied
e.g. in Refs.~\cite{Lukyanov2007,Lukyanov2009,Lukyanov2010,Lukyanov2013} for the analysis of elastic scattering data  of light exotic nuclei.
We already described details about the optical potential calculations in Refs.~\cite{Cow12,SCZ14}. Here we give, briefly, the main equations.

In  general, the potentials contain volume $V$ and spin-orbit $V_{\rm{SO}}$ parts,
which are both complex and expressed as
\begin{equation}\label{v}
U(\textbf{r})\,=\,V(\textbf{r}) \,+\,V_{\rm{SO}}(\textbf{r})\, {\bf L}\cdot {\bf S},
\end{equation}
where $\textbf{r}$ the radius-vector  connecting centers of the interacting nuclei, $\textbf{L}$ the angular momentum, and $\textbf{S}$ the intrinsic
spin of the projectile. In the $^4$He case, $\textbf{S}$=0 and the spin-orbit term
falls away.

We treat  the volume part of the optical potentials in the initial and the exit channels
on the same footing by application of the hybrid nucleus-nucleus optical potential.


The hybrid nucleus-nucleus optical potential \cite{Lukyanov2004} has real and imaginary parts:
\begin{equation}\label{eq:1}
U(\textbf{r})\, = \, N^R V^{\rm{DF}}(\textbf{r})\, + \, iN^I W(\textbf{r}).
\end{equation}
 The
parameters $N^R$ and $N^I$  correct the strength of the
microscopically calculated real $V^{\rm{DF}}$ and imaginary $W$ constituents of the whole potential.
They are usually adjusted comparing calculations of elastic cross sections to experimental data in the corresponding channels.
The real part $V^{\rm{DF}}$ is a double-folding potential that
consists of direct and exchange components:
\begin{equation}\label{eq:2}
V^{\rm{DF}}(\textbf{r})= V^D(\textbf{r}) + V^{\rm{EX}}(\textbf{r}),
\end{equation}
with
\begin{equation}\label{eq:3}
V^D(\textbf{r}) = \int d\textbf{r}_p \, d\textbf{r}_t \, {\rho}_p({\bf r}_p) {\rho}_t ({\bf
r}_t) v_{NN}^D(\textbf{s}).
\end{equation}
The exchange potential is
\begin{eqnarray}\label{eq:4}
&&\hspace{-1cm}V^{\rm{EX}}(\textbf{r}) \nonumber \\
&&\hspace{-0.8cm}= \int d\textbf{r}_p\, d \textbf{r}_t  \, {\rho}_p({\bf r}_p, {\bf r}_p+ {\bf s})  {\rho}_t({\bf r}_t, {\bf r}_t-{\bf s}) \nonumber \\
&&\hspace{2cm} \times \, v_{NN}^{\rm{EX}}(s)  \exp\left[ \frac{i{\bf K}(\textbf{r})\cdot \textbf{s}}{M}\right],
\end{eqnarray}
where ${\bf s}={\bf r}+{\bf r}_t-{\bf r}_p$ is the vector between
the projectile and target nucleons. The reduced mass coefficient
is $M~=~A_pA_t/(A_p+A_t)$, where $A_{p}$ and $A_{t}$ refer to
the projectile and target atomic mass numbers.
The radial part of the nucleus-nucleus momentum $K(r)$ is determined as follows:
\begin{equation}\label{eq:7}
K(r)=\left \{\frac{2Mm}{\hbar^2}\left[E-V^{\rm{DF}}(r)-V_c(r)\right ]\right \}^{1/2}.
\end{equation}
where $V_{c}$ is the Coulomb potential and $m$ is the nucleon mass.
The quantities $\rho_p({\bf r}_p)$ and $\rho_t({\bf r}_t)$ are
their density distributions,
$\rho_p({\bf r}_p, {\bf r}_p+{\bf s})$ and $\rho_t({\bf r}_t , {\bf r}_t-{\bf s})$
are the density matrices, which are approximated as in Ref.~\cite{Negele72}.
The CDM3Y6-type effective $N$-$N$ potentials $v_{NN}^{D}$  are based on the Paris $N$-$N$
potential determined in Ref. \cite{Khoa2000}.

For the initial channel calculations,  $\rho_{t}$ for $^{93}$Nb was taken as
the standard Fermi form,  with parameters from Ref. \cite{Pat03}.
In the exit channel a Fermi-form density with
parameters from Ref. \cite{elazab1985} was adopted for $^{90}$Zr, and the $^4$He density
from Ref. \cite{burov1977} was used.


The imaginary part of the optical potential $W(\textbf{r})$ in Eq.~(\ref{eq:1}) may have the
same form as its real counterpart  $V^{\rm{DF}}$, or  can be calculated separately within
the high-energy approximation  \cite{Glauber} as it was developed in Ref.~\cite{Lukyanov2004}.

The microscopic optical potential obtained in the high-energy approximation in the momentum space has the form:
\begin{eqnarray}\label{eq:8}
&&\hspace{-1.5cm}U^H_{\rm{opt}}(\textbf{r}) \,=\, -\frac{E}{k}{\bar\sigma}_{N} (i + {\bar\alpha}_{N})\frac{1}{(2\pi)^3} \;\;\;  \nonumber \\
&&\hspace{0.5cm} \times \int d\textbf{q } e^{\displaystyle{-i\bf q \cdot \bf r}}{\rho}_p(\textbf{q}){\rho}_t(\textbf{q})f_N(q) \,.
\end{eqnarray}
Here the $N$-$N$ total scattering cross section $\bar\sigma_{N}$ and the ratio of real to imaginary parts of the forward $N$-$N$
amplitude $\bar\alpha_{N}$ is averaged over the isospins of the projectile and target nuclei. They are parameterized as given
in Refs.~\cite{Charagi92,Shukla2001}.
The $N$-$N$ form factor is taken as $f_N(q)=\exp(-q^2\beta_N/2)$ with the slope parameter $\beta_N=0.219\,fm^2$ \cite{Alkhazov1978}.
In fact, we used only the imaginary part of Eq.~(\ref{eq:8}).


For the potential in the $p+^{93}$Nb channel,
the functions
$\rho_p(\textbf{r}_p)$ in Eqs.~\eqref{eq:3} and \eqref{eq:4} have to be excluded together with the elementary volumes $d\textbf{r}_p$.
Also, in  Eq.~\eqref{eq:8}, $\rho_p(\textbf{q})$
should not appear.

The shape of the analyzing power is rather sensitive to the spin-orbit part of the optical potential in the initial channel. Good agreement with the
experimental data was obtained by using for protons a Woods-Saxon shape of the real part of $V_{\rm{SO}}(r)$. We used the parameters  listed in Ref. \cite{BG}.

\begin{figure}[htb]
\includegraphics[scale=0.48]{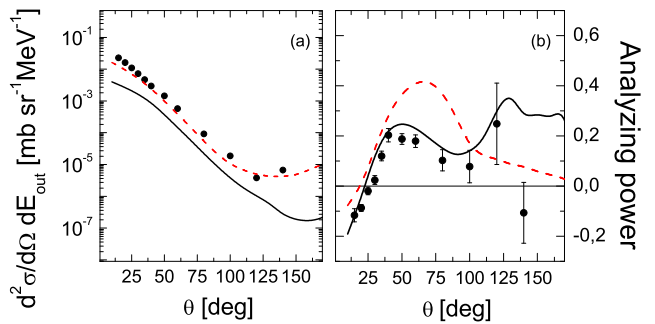}
\caption{(Color online) Double-differential cross section (a) and analyzing power (b) as a function of scattering
angle $\protect\theta $ for the $^{93}$Nb($p$,$\alpha $)
reaction at an incident energy of 100~MeV and an $\alpha $-particle emission energy of 86~MeV. Theoretical cross
section and analyzing power calculations for $N^I$=1 ($---$) and $N^I$=2 (solid line) in Eq.~(\ref{eq:1}) are compared with the experimental data.
Results for only the pickup component of the reaction mechanism are used to display the trends.}
\label{fig.2}
\end{figure}

The renormalization constants $N^R$ and $N^I$  in the initial channel cannot be defined independently since  there are no the respective data on elastic
scattering. Therefore they are kept equal to unity, while for the exit channel we need to adjust them to follow the emission-energy trend of the
experimental analyzing power data. Very good agreement with the experimental data for the highest emission energy E$_{\rm{out}}$=98~MeV can be obtained
if the values of $N^R$ and $N^I$ for the exit channel are kept equal to unity as well. For the rest of the outgoing energies we used the
values $N^R$=1 and $N^I$=2. Fig. \ref{fig.2} demonstrates the effect which the value of $N^I$ has on the differential cross section and
the analyzing power of the reaction at 86~MeV emission energy. Only the pickup component of the reaction mechanism is used to illustrate the sensitivity
to the renormalization of the imaginary potential.

The values found in our present investigation demonstrate that the  hybrid optical potential which we use is appropriate for
the energy range we consider, although they  are not consistent with those for the $^{93}$Nb($\vec{p}$,$\alpha $) reaction at
incident energy of 160 MeV \cite{SCZ14}. Clearly, a further theoretical analysis and complementary experimental studies for proper evaluation and interpretation are needed.

\subsection{Reaction Mechanism}

The mechanism of the direct  ($p,\alpha $) reaction has been discussed intensively over the years. For
example in Refs. \cite{TLU81, LLW82} the multistep direct reaction theory analysis of  ($\vec{p}$,$\alpha $)
reactions at 65~MeV and 72~MeV incident energies suggested that the reaction mechanism should be a pickup of
a triton. In Ref. \cite{Gadioli84} the authors show that calculations assuming pickup of a triton and knockout
of an $\alpha $ particle equally well fit the angular distribution and the analyzing power of $^{90,92}$Zr($p$,$\alpha)$
reaction to the ground state and the first few excited states, while in Ref. \cite{Bonetti89} the knockout mechanism is preferred
for describing transitions to the continuum. In Ref. \cite{SCZ14} we considered proton induced $\alpha$-particle emission at 160~MeV
incident energy and a wide range of emission energies. The existence of experimental data for just forward angles did not allow us
to make decisive conclusion about the reaction mechanism, but the calculations assuming that the  ejectile  originates from an
$\alpha$-cluster knockout in the final stage reproduce roughly the angular distributions of the measured
cross section and analyzing power as a function of  $\alpha$-particle emission energy.

The history of this debate suggests that it would be wise to consider DWBA calculations for both reaction mechanisms.
The theoretical results may then be compared with experimental data of the differential cross section and the analyzing power where the first step process is expected to dominate.
At an incident energy of 100 MeV
this is true at the highest $\alpha $-particle emission energy of 98~MeV that is available. Numerically the difference between both types of calculations lies in the form factor,
the incoming and the outgoing distorted waves are calculated using the same optical model potentials for protons
and $\alpha$ particles respectively. The proton-triton binding potential has a  Woods-Saxon shape with
geometrical parameter $r_0=1.87$~fm and $a=0.7$~fm, whereas to calculate the $\alpha $-particle form factor we use the generally accepted geometrical parameter values  of $r_0=1.25$~fm and $a=0.65$~fm.

As is seen in Fig.~\ref{fig.3}, panel (a), the theoretical double-differential cross sections have rather different shapes for a knockout or pickup reaction mechanism. Whereas the pickup
cross section can be scaled to fit the forward angles, the knockout cross section reproduces the experimental data very well at larger angles.
The sum of the cross sections originating from both reaction mechanisms is required for a good fit to the complete set of experimental data over the whole range of scattering angles.
The scaling factors, which are needed to fit the experimental differential cross sections at 98~MeV emission energy,  are kept unchanged
for the rest of the calculations at other outgoing energies.

Panel (f) of  Fig.~\ref{fig.3} shows the analyzing power as a function of the scattering angle for pickup (pu) and knockout (ko) reaction mechanisms.
The solid line in the figure represents the sum of both distributions, defined as follows:

\begin{equation}
A_{\rm{total}}=\frac{A_{\rm{pu}}\left( \frac{d^{2}\sigma }{d\Omega dE}\right)
^{\rm{pu}}\!\!+A_{\rm{ko}}\left( \frac{d^{2}\sigma }{d\Omega dE}\right)
^{\rm{ko}}}{\left( \frac{d^{2}\sigma }{d\Omega dE}\right)
^{\rm{pu}}\!\!+\left( \frac{d^{2}\sigma }{d\Omega dE}\right)
^{\rm{ko}}}\;\;,
\end{equation}
\noindent where the subscripts and superscripts refer to either pickup or knockout, as the abbreviated notations imply.

The analyzing power distribution calculated assuming only pickup reproduces the complete set of experimental data reasonably well, but inclusion of the knockout
contribution is crucial. Clearly, both reaction mechanisms play an important role in the theoretical description of the $^{93}$Nb($p$,$\alpha $)
pre-equilibrium reaction. Their different contributions under various kinematical conditions are very noticeable.

Many superficially equivalent, yet inherently very different sets of optical potentials which all successfully
reproduce elastic scattering are available for
generating distorted waves in DWBA calculations of pre-equilibrium reactions. However, as was pointed out
in for example Ref. \cite{Cow00}, because of known problems, caution is advisable
in the choice of a specific set for pre-equilibrium reactions.
For application to pre-equilibrium reactions, the optical potentials must be valid over a large range of
incident and emission energies.  Furthermore, to be generally useful, a wide variety of target nuclei and different
ejectiles need to be covered by a single optical potential set.
The possible simultaneous importance of all these characteristics is suggested by
the observation (see for example Refs. \cite{Cow96,Cow07}) that experimental pre-equilibrium cross
section distributions of all types appear to follow closely the simple phenomenological systematics of Kalbach \cite{Kal}.
Consequently, implementation
of a single truly global optical potential which satisfy all requirements
is highly desirable.

In an earlier ($p$,$\alpha $) investigation \cite{Cow96} we used global phenomenological
distorting potentials, but these have now been abandoned in favor of a folding procedure. The reason
is that the former potentials are extracted independently for the projectile and for the ejectile from
elastic scattering, therefore the relationship between the two sets is unknown. In other words, it is
simply not clear whether they form a matched pair, as would be desirable. Furthermore, this relationship cannot
be readily checked for a reaction into the continuum, as for transfer to a discrete final
state. Consequently, in addition to those properties already discussed at
the beginning of this subsection, an optimal optical potential should for our present needs
offer a good
description of elastic scattering for proton projectiles as well
as $\alpha $-particles (and also for $^3$He to link up to our ongoing
two-nucleon transfer studies).
A folding procedure comes close to satisfying all the criteria,
but unfortunately as we have seen, at the cost of two free parameters, of which
one needs to be adjusted in this work. Nevertheless, as was already mentioned earlier, this
specific type of folding potential
has been successfully employed in the past for pre-equilibrium reactions and, of course, for elastic scattering.

\subsection{Influence of momentum mismatch between entrance- and exit channels}
Proton-induced multi-nucleon transfer reactions suffer from severe momentum mismatch
between the incident- and outgoing channels, which becomes progressively worse with
increasing projectile energy. For this reason it is generally accepted that reactions such
as ($p$,$t$) and ($p$,$\alpha $) cannot, even at low incident energies, provide reliable
spectroscopic information.

At the incident energies investigated in this work the momentum mismatch is in the
range of 400 to 600 MeV/c. At those momenta the asymptotic tail of the bound-state
wave function has decreased by many orders of magnitude from its maximum. Within
normal uncertainties this means that, for all practical purposes, the true value of the
bound-state wave function is unknown at the specific momentum range for which the
cross section is sensitive. Consequently, under those conditions extremely small
errors on the bound state influence cross-section values calculated in DWBA
enormously, rendering predicted absolute values meaningless. Fortunately this
difficulty does not influence the shape of the angular distribution
appreciably (see for example Ref. \cite{JJ}), therefore analyzing
power, which is a ratio of cross sections, is unaffected by the problem.

To address the problem in this work, we simply normalize our theoretical DWBA
cross-section values to the experimental pre-equilibrium angular distributions where
the reaction mechanism is likely to be purely of a direct one-step nature – in other
words at the highest emission energies. The same normalization is used at lower emission
energies where multistep contributions, which may be associated with lower initial nucleon driving energies,
become relevant. Clearly our procedure only partially
solves the problem towards lower emission energies, because it is not known to what extent
the adopted bound state reproduces the true trend of the bound-state wave function
towards lower incident energy correctly.

A recent investigation \cite{JJ} of the $^{58}$Ni($p$,$^{3}$He)$^{56}$Co reaction to discrete final
states, which allows an accurate extraction of the trend with incident energy, suggests that our
simplistic procedure could easily lead to a cross-section discrepancy as large as observed in
the present study at an incident energy of 100 MeV, as will be quantified later. This will be
discussed later.



\section{Results and discussion}

\begin{figure}[htb]
\includegraphics[scale=0.48]{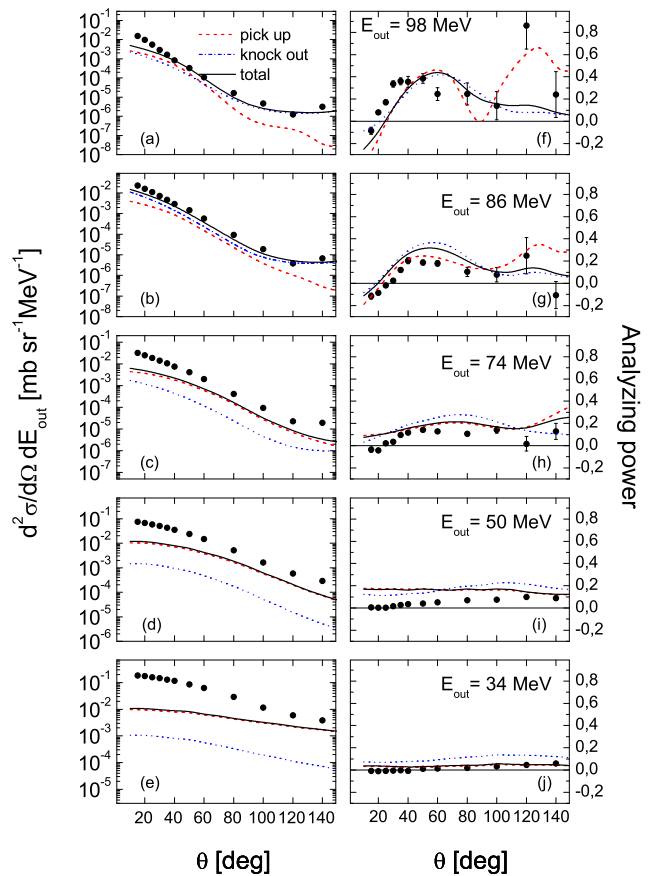}
\caption{(Color online) Double-differential cross sections (a)-(e) and analyzing power (f)-(j) as a
function of scattering angle $\protect\theta $ for the $^{93}$Nb($p$,$\alpha $)
reaction at an incident energy of 100~MeV and various $\alpha $-particle emission
energies $E_{\rm{out}}$ as indicated. Theoretical cross section calculations for
pickup ($---$) and knockout
($- \cdot - $)
are shown, with the sums of both reaction mechanisms plotted
as continuous curves. The experimental analyzing power distributions are
compared with theoretical calculations for pickup ($---$), knockout ($- \cdot - $) and the sum of both reaction mechanisms (solid lines).}
\label{fig.3}
\end{figure}

We summarize the results of our calculations in Fig.~\ref{fig.3} where the double-differential cross section and analyzing power angular distributions for the $^{93}$Nb($p$,$\alpha$) reaction
at an incident energy of 100~MeV  for various outgoing energies of the $\alpha$ particles are displayed.

Experimental data are available for outgoing energies starting from 98~MeV (with 106~MeV as a kinematic limit due to a positive $Q$-value of the reaction of 6.4~MeV) down to 34~MeV.
We have chosen the ones shown in the figure because they are  representative of the contribution of both reaction mechanisms to the total differential cross section and
analyzing power, respectively. The theoretical results are in good agreement with the experimental quantities, although we have to keep in mind that we have fitted
some of the input ingredients of the theory, as implied in Sec. III.

All the theoretical double-differential cross section distributions were  normalized with the factors extracted from  the angular distributions at an emission
energy $E_{\rm{out}}$ of  98~MeV, for which the one-step reaction dominates,
as explained in Sec.~IIIE. Although the fitting procedure is
based on theoretical considerations, it is still somewhat arbitrary. As we explained in Ref. \cite{SCZ14}, experimental uncertainties in, for example,
the emission energy calibration would result in a systematic error in the measured cross section which rapidly gets worse towards the top end of emission energies.
The reason is that the energy distribution of the cross section as function of emission energy drops very rapidly to zero as the kinematic limit is approached,
whereas it varies considerably more slowly at lower emission energies. Our cross section data, for an incident energy of 100 MeV, at the highest emission energy is already in an energy range
where a rapid variation occurs.  This, combined with the experimental uncertainty in emission energy, could affect the reliability of the normalization procedure.

A noticeable trend displayed in Fig.~\ref{fig.3} is that the theoretical cross sections underestimate the
experimental values increasingly towards lower emission energies. As discussed in Sec.~IIIF
this could perhaps be a normal phenomenon caused by inadequacies in the way that the effect
of momentum mismatch is compensated. However, other possible causes should also be
considered. It is significant that a similar trend with emission energy was observed even more
severely in our other ($p$,$\alpha$) studies at higher incident energy \cite{Cow96,SCZ14}. In
one of these investigations \cite{SCZ14} the same folding procedure as in this work was used
to generate distorting potentials, whereas in the other study \cite{Cow96} normal global phenomenological
optical potentials were employed. The similarity of the discrepancy encountered in the two cases would seem to
rule against the specific choice of optical potential in this work as a possible cause
of the difficulty. A different issue, unrelated to the optical potential, is that $\alpha $-particle evaporation from a compound
nucleus
could in principle contribute to the continuum yield, thus explaining the observed under-prediction. However, explicit calculation
shows that such mechanism contributes only a fraction of a percent \cite{TENDL} to the total angle-integrated cross section at
the lowest
emission energy of Fig.~\ref{fig.3}. An isotropic
distribution for the evaporation component still implies that it contributes less than 1\% to the yield even at the most backward angle.
A similar negligible contribution is present at an incident
energy of 65 MeV and at an emission energy of 37 MeV, which will be explored later. In other
words, $\alpha $-particle evaporation
from compound-nuclear decay definitely does not distort any of the angular distributions displayed in this work.

As was mentioned earlier, and also as was pointed out in Ref. \cite{SCZ14},
because the analyzing power consists essentially of a ratio of cross sections,
it would not be appreciably affected by most of the putative causes
of a cross section problem. It is significant that the experimental analyzing-power angular
distributions are reproduced well by the
theory over the whole range of emission energies explored. Because we overwhelmingly base
our conclusions regarding the reaction mechanism on features observed in analyzing power distributions,
we do not consider the cross section under-prediction to be a serious concern.


As is also shown in Fig.~\ref{fig.3}, the differential cross sections of the knockout reaction mechanism decrease faster
than those for pickup towards lower emission energies. Therefore, on average the
total differential cross section is dominated by the pickup contribution at an incident energy of 100 MeV.

\begin{figure}[htb]
\includegraphics[scale=0.43]{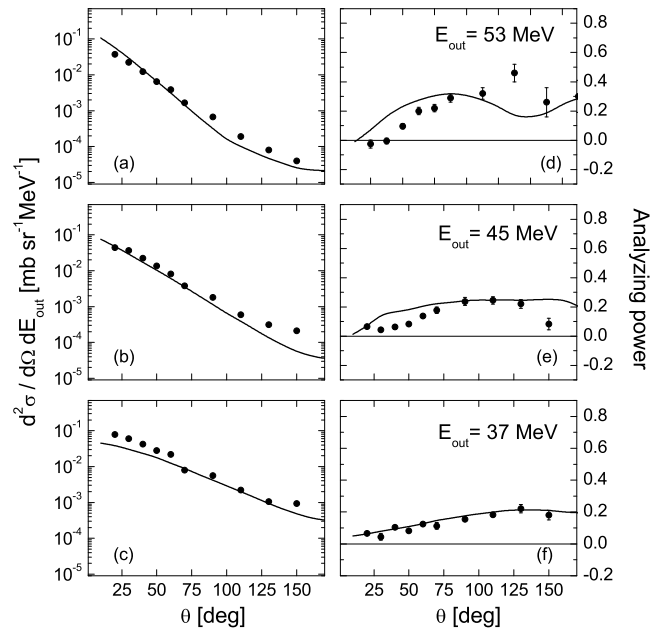}
\caption{Double-differential cross sections (a)-(c) and analyzing power (d)-(f) as a function of scattering angle $\protect\theta $ for
the $^{93}$Nb($p$,$\alpha $) reaction at an incident energy of 65~MeV and various $\alpha $-particle emission energies $E_{\rm{out}}$ as
indicated. Theoretical calculations for a knockout reaction mechanism (solid line)
are compared with the experimental data by Sakai \textit{et al.}  \cite{Sakai80}.}
\label{fig.4}
\end{figure}

To  extend the study of the  $^{93}$Nb($p$,$\alpha$)  reaction to a  lower incident energy we re-examined the experimental data by Sakai \textit{et al.} \cite{Sakai80},
where the differential cross section and the analyzing power distributions of the continuum spectra for various target nuclei including $^{93}$Nb were measured for
65 MeV polarized protons in a wide range of excitation energies and angles.

Our further investigation of the  $^{93}$Nb($p$,$\alpha$) reaction at 65~MeV incident proton energy and at three outgoing energies
followed the same procedures as those described in Sec.~III.
It turned out that for this incident energy the knockout mechanism is sufficient to describe the available experimental data. Not only does a pickup mechanism
give inferior agreement with the experimental angular distributions, but any combination of pickup and knockout fails to achieve better results than knockout by itself.

The comparison
of the experimental and theoretical double-differential cross section and analyzing power is shown in Fig.~\ref{fig.4}. First of all we should
point out that the theoretical calculations reproduce  the shape of the differential cross section at the largest outgoing energy
of 53~MeV very well. Once fitted at this emission energy the magnitudes of the differential cross section are in very good agreement with the experimental data
at lower emission energies as well. We may speculate that this is because we treat all important intranuclear processes properly, but we
should also keep in mind that we determine the scaling factor at an emission energy as much as 18~MeV lower than the kinematic limit,
where the first-step direct knockout is no longer the only kinematically allowed process. Furthermore, we also explore a very limited emission-energy range of only 16 MeV, as provided by the available experimental data.
Nevertheless, the magnitude and shape of the cross section, as well as the shape of the  analyzing power distributions
are  reproduced remarkably well at all emission energies which are available. Consequently we can confidently claim that at 65~MeV incident energy the  $^{93}$Nb($p$,$\alpha$) reaction is described
mainly by a knockout reaction mechanism.

These data of Sakai \textit{et al.} \cite{Sakai80} at an incident energy of 65 MeV have also been investigated by Tamura \textit{et al.} \cite{TLU81}. It is beyond the scope of the present work
to discuss details of the calculations of Ref. \cite{TLU81}
in such a way that a proper comparison with our results is meaningful. Nevertheless, it is noteworthy that the analyzing power angular distributions are reproduced
considerably better in our work.

\begin{figure}[htb]
\includegraphics[scale=0.43]{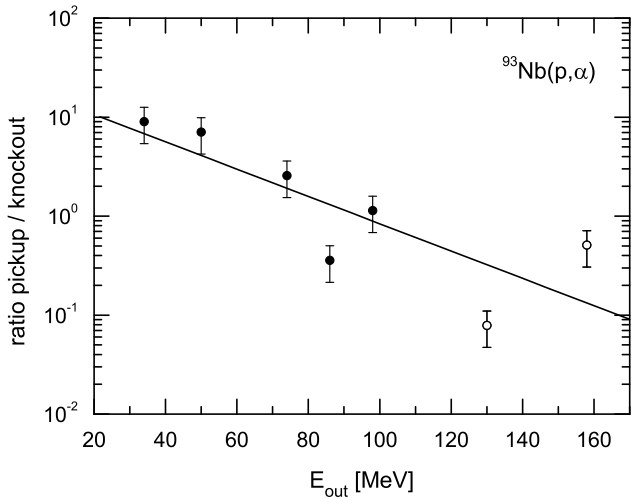}
\caption{Ratio of cross sections for pickup and knockout as a function of $\alpha $-particle emission energy $E_{\rm{out}}$. Solid circles represent results for an incident energy of 100 MeV, and open circles are for 160 MeV
from previous work \cite{SCZ14}. Error bars are only rough estimates, and the curve is an exponential fit to the data.}
\label{fig.5}
\end{figure}


An interesting feature of the theoretical cross section angular distributions in Fig. \ref{fig.3}, for an incident energy of 100 MeV, is that the magnitudes for knockout drop off more rapidly with decreasing emission energy
than those of pickup. This is plotted in Fig. \ref{fig.5} as the ratio of pickup to knockout as a function of emission energy. Of course the shapes of the two distributions
differ somewhat, therefore the most forward angle at which experimental data were measured was arbitrarily chosen for the comparison. Clearly the observed trend would be only
slightly influenced by this specific choice. We find that for an incident energy of 100 MeV pickup becomes more important with decreasing emission energy. Of course, emission energy is
simply related to incident energy, but due to the changing contributions from different reaction steps, one should not simply infer from this that pickup should be even more important at 65 MeV than at 100 MeV
incident energy. In fact, we find the opposite. As shown in Fig. \ref{fig.5}, the result from our previous study \cite{SCZ14} at 160 MeV does indeed seem to support such a naive incident-energy dependence.
However this result may be spurious if we consider the large systematic uncertainty in the extracted ratio in that case. At 100 MeV incident energy, not only is the systematic uncertainty
much smaller, but the trend is very reliable.

Our current optical potentials do not comprise a surface imaginary component. This neglect could in principle become an increasing problem
towards lower incident energies. Evidence of the importance
of such deficiency would presumably have revealed itself at 65 MeV as a worsening agreement between the theoretical and
experimental distributions, or perhaps as an unfavorable comparison of our results with those of Tamura \textit{et al.} \cite{TLU81}, in whose work
surface effects are included. We reassuringly find exactly the opposite in both cases. We should mention that, as a whole, our various
pre-equilibrium studies published elsewhere \cite{Cow96,Cow97,Cow00,Cow12,Cow07,SCZ14} do not seem to show a qualitative
sensitivity to the exact choice of any reasonable optical potential. This may be because the multistep character of the reaction
mechanism puts a powerful stamp on the features of the experimental observables.


\section{Summary and conclusions}
Cross section and analyzing-power angular distributions for the reaction $^{93}$Nb($\vec{p}$,$\alpha $) at a projectile energy of 100~MeV and various $\alpha $-particle emission
energies from 98 MeV down to 34 MeV were presented.
The experimental angular distributions were compared with the predictions of a statistical multistep direct emission. Reasonable agreement was found between theoretical and experimental results if
both knockout as well as pickup are included as mechanisms leading to the final emission of $\alpha $ particles.

The same theoretical analysis was extended to existing experimental results \cite{Sakai80} for the same reaction at a lower projectile energy at 65 MeV. The predicted cross section and analyzing-power angular distributions
were again in very good agreement with the experimental data. However, at this incident energy, a strong preference was found for a knockout process. This finding is in agreement with our earlier work at an incident energy of 160 MeV
\cite{SCZ14}.

Evidently the reaction mechanism in the  $^{93}$Nb($\vec{p}$,$\alpha $) reaction changes from a dominant knockout process at 65 MeV incident energy, to a combination of pickup and knockout
participating at 100 MeV, and then back to only knockout being important at 160 MeV. The usual assumption is that a target such as $^{93}$Nb is representative of nuclei in general
as far as the pre-equilibrium ($p$,$\alpha $) reaction is concerned. However, the present conclusion regarding the change in the ratio of participating mechanisms for this target needs to be confirmed for other nuclear species.

As in other investigations of the ($p$,$\alpha $) reaction \cite{Cow96,SCZ14}, at 100-MeV incident energy it is found that the absolute cross section is increasingly under-predicted towards lower emission energies. Although
comparisons between experimental and theoretical analyzing power results suggest that this is not a serious concern, it would be advisable to investigate the lower-than-expected theoretical cross sections further. Based on the trend from
comparable ($p$,$^3$He) \cite{Cow97,Cow00,Cow07,Cow12} and ($p$,$\alpha $) \cite{Cow96,SCZ14} studies in the 100 MeV to 200 MeV incident-energy range, it is reasonable to speculate that the issue
which is encountered in the magnitude of the predicted cross section is mainly related to problems with proton-induced multi-nucleon transfer reactions in general. For example, it is well known that severe momentum mismatch in
($p$,$\alpha $) reactions to discrete final states makes it difficult, if not impossible, to extract spectroscopic information. This is caused by sensitivity of the cross section to the asymptotic region of the
bound state wave function, due to momentum mismatch, which is sampled by a ($p$,$\alpha $) reaction.

Clearly, it would be informative to explore the issues found in the present investigation for other targets. Experimental as well as further theoretical work should be invaluable.

\begin{acknowledgments}
Our appreciation is extended to G. F. Steyn and S. V. F\"{o}rtsch for providing
the numerical data at an incident energy of 100 MeV reported in this work.
The research of A.A.C. was funded by the National Research Foundation (NRF)
of South Africa. The studies of S.S.D. were partially supported by the SARFEN grant of the Bulgarian
Science Foundation. E.V.Z. and K.V.L were supported by the Russian Foundation for Basic Research (RFBR)
under Grants No. 12-01-00396a and No. 13-01-00060a. This financial support is gratefully acknowledged.
\end{acknowledgments}

\end{document}